\documentclass[aip,jcp,amsmath,amssymb,floatfix,reprint,citeautoscript,articletitle=false,noeprint]{revtex4-2}

\usepackage{amssymb}
\usepackage{amsmath}
\usepackage{graphicx}
\usepackage{dcolumn}
\usepackage{xcolor}
\usepackage{fancyhdr}
\usepackage[colorlinks,allcolors=black,citecolor=blue,urlcolor=blue]{hyperref}

\begin{document}

\title{When Is Nanoconfined Water Different From Interfacial Water?}

\author{Xavier R. Advincula}
\email{xr223@cam.ac.uk}
\affiliation{Yusuf Hamied Department of Chemistry, University of Cambridge, Lensfield Road, Cambridge, CB2 1EW, UK}
\affiliation{Cavendish Laboratory, Department of Physics, University of Cambridge, Cambridge, CB3 0HE, UK}
\affiliation{Lennard-Jones Centre, University of Cambridge, Trinity Ln, Cambridge, CB2 1TN, UK}

\author{Christoph Schran}
\email{cs2121@cam.ac.uk}
\affiliation{Cavendish Laboratory, Department of Physics, University of Cambridge, Cambridge, CB3 0HE, UK}
\affiliation{Lennard-Jones Centre, University of Cambridge, Trinity Ln, Cambridge, CB2 1TN, UK}

\author{Angelos Michaelides}
\email{am452@cam.ac.uk}
\affiliation{Yusuf Hamied Department of Chemistry, University of Cambridge, Lensfield Road, Cambridge, CB2 1EW, UK}
\affiliation{Lennard-Jones Centre, University of Cambridge, Trinity Ln, Cambridge, CB2 1TN, UK}

\begin{abstract} 
Water behaves very differently at surfaces and under extreme confinement, but the boundary between these two regimes has remained unclear.
Despite evidence that interfacial effects persist under sub-nanometre confinement, the molecular-scale behaviour and its evolution with slit width remain unclear.
Here, we use machine-learning molecular dynamics with first-principles accuracy to probe water at graphene surfaces across slit widths ranging from the open-interface limit to angstrom-scale confinement.
We find that water undergoes a sharp structural transition: when three or more water layers fit between the walls, the structure of the graphene-water interface is effectively indistinguishable from that in an open system, with density layering, hydrogen bonding, and orientational ordering retaining interfacial character.
Below this threshold, however, angstrom-scale confinement strongly reorganises the liquid, producing enhanced ordering, a restructured hydrogen-bond network, and modified orientational motifs.
These results establish a molecular-level picture that clearly separates interfacial behaviour from genuine nanoconfinement and provide guidance for predicting and controlling the structure of water in nanoscale solid–liquid environments.

\end{abstract}
\maketitle


\section*{Introduction}

Civilisations have often been described by the materials they learned to use: the Stone, Bronze, and Iron Ages, and more recently, the so-called Silicon Age.
As we move through the 21st century, another material is coming to the forefront.
Water is our planet’s most essential resource, and the second half of this century will require us to understand and control it in far more sophisticated ways.
Clean water is crucial for human survival, and water will also play an increasingly important role as an energy carrier in efforts to tackle climate change.
It is in this sense that we might start talking about a new Water Age.
A key part of this emerging Water Age will involve technologies that rely on water at interfaces and under nanoconfinement.
Improving membranes for filtration and desalination, using water in electrochemical energy technologies, or harvesting blue energy all depend on how water behaves near surfaces and in narrow pores.
We know that interfacial and nanoconfined water show many unusual and fascinating properties.
Yet, despite decades of work, the line between ``interfacial'' and ``nanoconfined'' water is still not sharply defined.
This article looks at where that distinction really lies.

When water is confined at the nanoscale, its structural and dynamical properties depart markedly from those of the bulk, influencing processes in nanofluidics~\cite{lyderic_2010_rev, Radenovic_2024}, electrochemistry~\cite{Lozada-Hidalgo2018, electrochem_conf, Gomes2024}, energy conversion~\cite{siria_energy_2017, Zhang2021_energy_conv}, and biological function~\cite{Murata2000, gravelle_2013}.
Experiments and simulations have shown that molecular confinement can alter dielectric screening~\cite{fumagalli_dielectric_2018, barragan_dielectric_2020, prl_helene_2021, chris_dufils_2024, Wang2025_fumagalli}, induce layering or crystallisation~\cite{tanaka_2008, Han2010, koga_nanopores_2015, algara-siller_square_2015, vk_chris_2022, vk_chris_2022_correction, jiang_rich_2024}, and modify water transport~\cite{holt_2006, Duan2010, Hu2014, Ma2015, radha_b_2016, noy_2017_science, Keerthi2018, geim_2019, kavokine_friction_2022, Li2023_nano, Coquinot2025_flow}.
These effects have been reported across slit pores~\cite{galli_canonical_2008, Han2010, sergi_nanoconf_2019, vk_chris_2022, vk_chris_2022_correction}, nanotubes~\cite{hummer_2003,holt_2006, galli_canonical_2008, Ma2015, Agrawal2017, noy_2017_science, flt_nanotubes_2022, Li2023_nano}, and atomically thin membranes~\cite{Hu2014, radha_b_2016, Keerthi2018, geim_2019}.
In these settings, confinement can impose strong layering and orientational ordering, and may even drive structural transitions absent in the bulk~\cite{franzese_2021, lyderic_crossroads_2023, Trushin2025}.

At the same time, even open solid–liquid interfaces impart significant structural and dynamical perturbations~\cite{wat_inter_2016, steve_air_graphene_2017, ohto_2019, wang_ange_2024, gra_wat_acid_2025, ions_gra_advin_2025}.
Graphene–water, metal–water, and other hydrophobic or hydrophilic interfaces exhibit truncated hydrogen-bond networks, orientational ordering, and suppressed capillary fluctuations~\cite{wat_inter_2016, steve_air_graphene_2017, ohto_2019, wang_ange_2024}.
Because these perturbations can extend multiple molecular layers into the liquid, they raise an important and unresolved question: How wide must a channel be before its behaviour differs from that of two overlapping interfacial layers?

While experimental work has reported confinement-like signatures persisting up to nanometric scales~\cite{fayer_2007, lyderic_2010_rev, Duan2010, fumagalli_dielectric_2018, gao_2024, Wang2025_fumagalli}, many simulations predict a much shorter decay length of only a few molecular layers~\cite{chialvo_2016, sergi_nanoconf_2019, martina_dominik_2022, chris_dufils_2024, netz_thz_ir_2024}.
This discrepancy reflects a broader conceptual gap\cite{know_gap_2019}: although interfacial and confined water have each been extensively studied, a continuous, molecular-level comparison between them is still missing.

Here we address this question by combining first–principles–level accuracy with long-timescale sampling enabled by machine learning potentials (MLPs).
Using the MLP developed in this work, we simulate water at graphene surfaces across slit widths ranging from the open-interface limit to angstrom-scale confinement, and compare these directly with graphene–water–vacuum systems.
The use of an MLP allows us to access system sizes and trajectory lengths that are not feasible with direct \textit{ab initio} molecular dynamics while retaining the accuracy of the underlying electronic-structure method~\cite{jorg_2016, sciadv_2017_ceriotti, montero_2024, Thiemann_2025_tutorial}.
This approach enables a consistent, molecular-level comparison of how confinement modifies density layering, hydrogen bonding, and orientational ordering, three microscopic descriptors central to interfacial water structure.

Our results reveal a sharp transition that clearly distinguishes interfacial behaviour from genuine nanoconfinement.
When more than three molecular layers fit between graphene sheets, the water structure adjacent to each surface is essentially indistinguishable from that at an open graphene–water interface: the first-layer density peak, the hydrogen-bond environment, and the orientational distributions all converge onto the interfacial limit.
Below this threshold, however, confinement leads to a marked reorganisation of the liquid.
The structural, hydrogen-bonding, and orientational signatures deviate strongly from their interfacial counterparts, marking the onset of a distinct nanoconfined regime.
Beyond identifying this crossover, the separation we find provides a molecular-scale framework for interpreting experiments on nanoscale water.

\begin{figure*}[htp!]
    \centering
    \includegraphics[width=\linewidth]{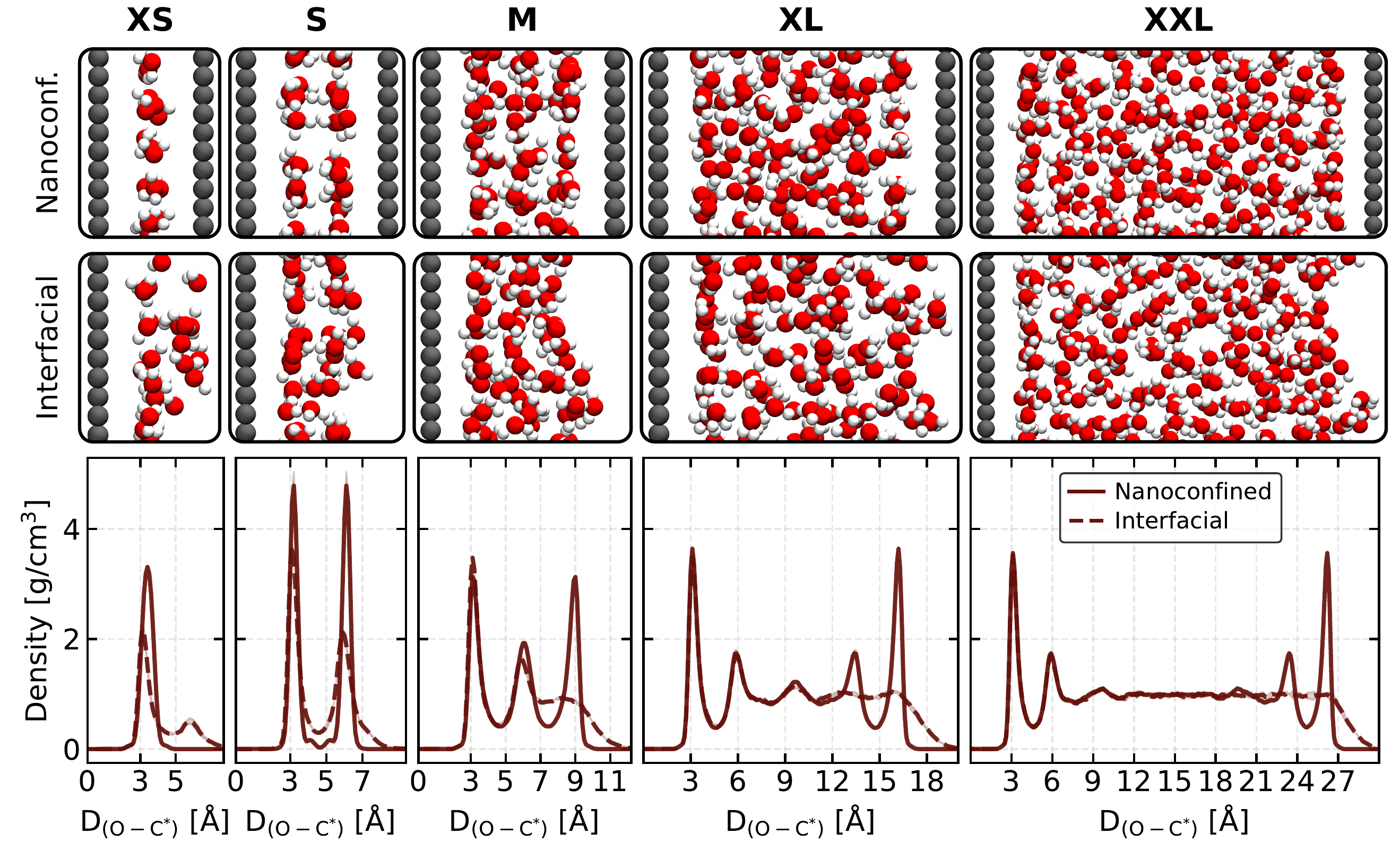}
    \caption{Representative snapshots and corresponding water density profiles for nanoconfined slit-pore systems and interfacial graphene–water–vacuum systems.
    All profiles are referenced to a common graphene reference plane, and the distance to this plane is denoted $D_{(\text{O}-\text{C}^{*})}$, allowing direct, width-independent comparison between nanoconfined and interfacial environments.
    }
    \label{fig:fig1}
\end{figure*}

\section*{Results}

\subsection*{The Liquid Structuring in Strongly Confined Water Is Different From Weakly Confined Water}

We examined five graphene slit-pore systems in which water is confined between parallel graphene sheets, spanning pore widths from roughly 7 to 30~{\AA}.
Following convention in this field, we refer to the narrow pores ($<\sim$10~{\AA}) as offering ``strong'' confinement and the wider pores  ($>\sim$10~{\AA}) as providing ``weak'' confinement.
The particular geometries used here follow those introduced in Ref.~\citenum{sergi_nanoconf_2019}, adopting the same lateral dimensions and water packing.
For clarity, we refer to the pores as extra-small (XS, 6.91~{\AA}), small (S, 9.41~{\AA}), medium (M, 12.20~{\AA}), extra-large (XL, 19.41~{\AA}), and extra-extra-large (XXL, 29.41~{\AA}), as illustrated in Figure~\ref{fig:fig1}.
Because the XXL slit lies beyond the pore sizes treated in Ref.~\citenum{sergi_nanoconf_2019}, we constructed it by extrapolating the XL system’s density so that the density approaches a bulk-like value at the centre of the slit.
In addition to the graphene–water–graphene pores, we constructed analogous systems in which one of the confining graphene sheets is removed, producing graphene–water–vacuum systems (hereafter referred to as interfacial systems).
This setup allows us to directly compare the graphene–water interface in the nanoconfined pores with that in an open interfacial environment.
Because removing one sheet exposes the liquid to vacuum, these interfacial systems also provide a useful point of reference for contrasting solid–liquid structuring with the behaviour at a free (air–water–like) surface.
To model these systems, we developed an MLP based on the revPBE-D3(0) density functional, extending the training data from our previous work~\cite{gra_wat_acid_2025} and employing the MACE architecture~\cite{mace_canonical} (see Methods for details).

\begin{figure*}[htp!]
    \centering
    \includegraphics[width=\linewidth]{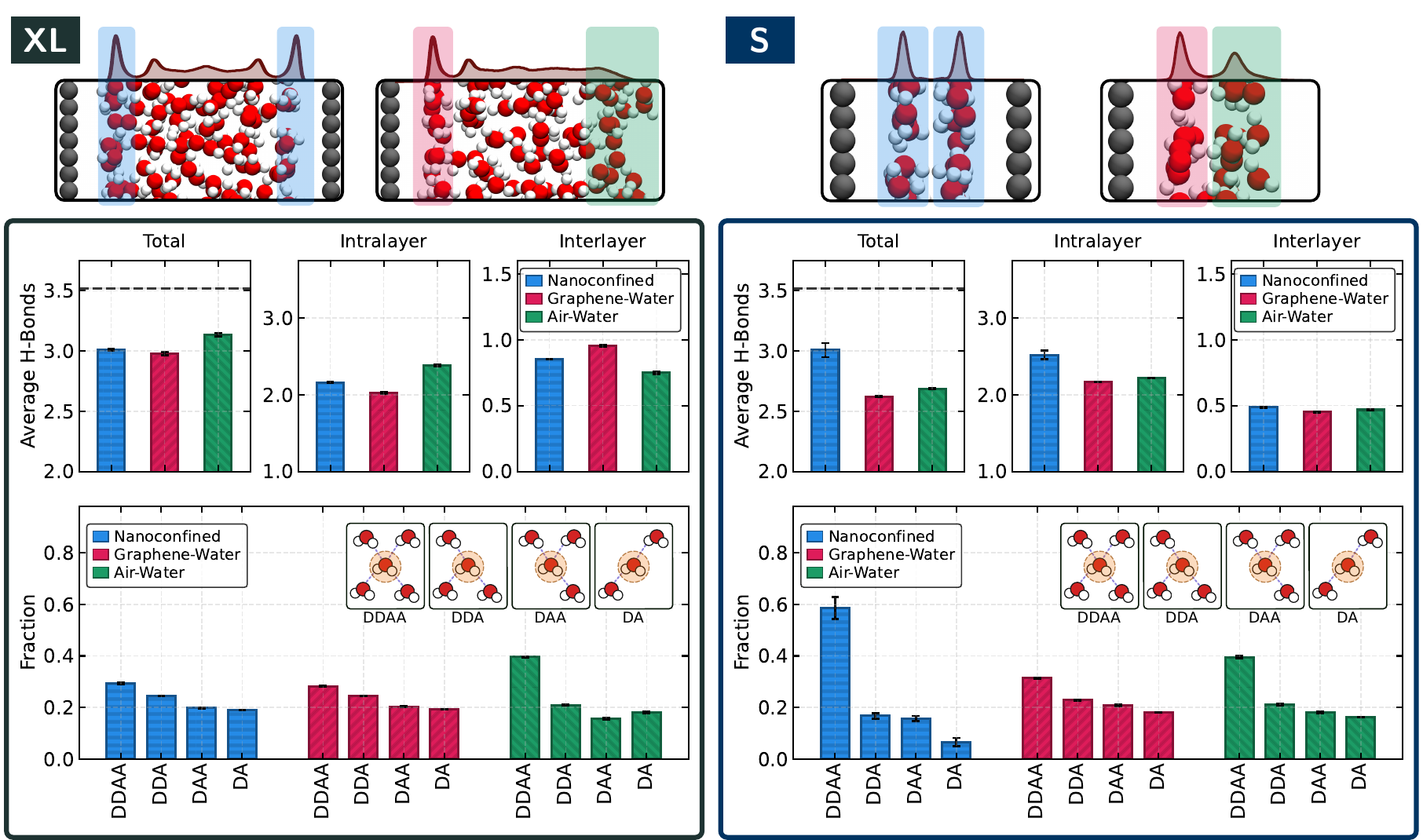}
    \caption{Confinement-induced differences in the interfacial hydrogen-bond network.
    Representative snapshots illustrate the systems and the definition of the interfacial regions.
    The panels show the average number of hydrogen bonds, classified as total, intralayer, and interlayer, together with the distribution of donor–acceptor bonding motifs (D = donor, A = acceptor).
    In the average–hydrogen-bond plots, the horizontal dashed line marks the bulk-water value.
    The left panel shows results for the XL systems and the right panel for the S systems.
    }
    \label{fig:fig2}
\end{figure*}

From Figure~\ref{fig:fig1}, we observe that both the nanoconfined and interfacial systems show pronounced density stratification near graphene, consistent with previous studies of water at solid surfaces~\cite{galli_canonical_2008, tocci_friction_2014, sergi_nanoconf_2019, gra_wat_sfg_charged_paesani_2025}.
Interestingly, this behaviour differs substantially from the much weaker layering at the air–water interface, reflecting the stronger structural imprint of the solid substrate~\cite{wc_2010, pezzotti_2017, steve_air_graphene_2017, litman_2024_nature, ions_gra_advin_2025}.
This contrast is also apparent in the structural snapshots, particularly for the M system: nanoconfined water forms clear, well-defined layers, whereas the graphene–water–vacuum configuration exhibits a visibly less defined interface shaped by capillary fluctuations.
This highlights that graphene’s rigidity suppresses such fluctuations, producing a more ordered interfacial environment than the free surface.

By comparing the density peaks adjacent to graphene in the nanoconfined and interfacial systems, we find a clear distinction between the wider and narrower pores.
For the wider pores (XL and XXL), the separation is greater than one nanometre, which allows several molecular layers to form.
In this ``weak-confinement'' regime, the first-layer peak is essentially identical in both environments, reaching roughly 3 g/cm$^{3}$.
For the M system, which accommodates three layers, the interfacial peak is already very close to the open-interface value, although small deviations remain.
We therefore regard this pore as lying at the boundary between interfacial-like and confined behaviour.
Taken together, these trends indicate that once more than three layers can form, the influence of the opposite confining surface becomes negligible and the graphene–water interface behaves effectively the same in both setups.
In contrast, the narrowest pores (XS and S) are less than a nanometre wide and accommodate fewer than two molecular layers.
These systems fall into the ``strong-confinement'' regime, where the peak positions and intensities differ markedly from those in the interfacial systems.
Here, the limited available space strongly restricts the structure of the interfacial film, leading to a sharper and more compressed first layer in the nanoconfined case.
Angstrom-scale confinement in this regime amplifies the ordering imposed by the solid surface, whereas this influence dissipates and eventually vanishes once additional layers can form.
Interestingly, in the XS system, removing one graphene sheet results in the formation of small water clusters rather than a continuous film, which exhibits a bimodal density profile with a significantly reduced overall density.
This outcome highlights how extreme confinement drives the system into a qualitatively distinct interfacial regime.
In contrast, these effects diminish and ultimately disappear as the confinement is relaxed and further layers can form.

\subsection*{Breakdown of Interfacial-Like Hydrogen Bonding Under Strong Confinement}

\begin{figure*}[htp!]
    \centering
    \includegraphics[width=0.85\linewidth]{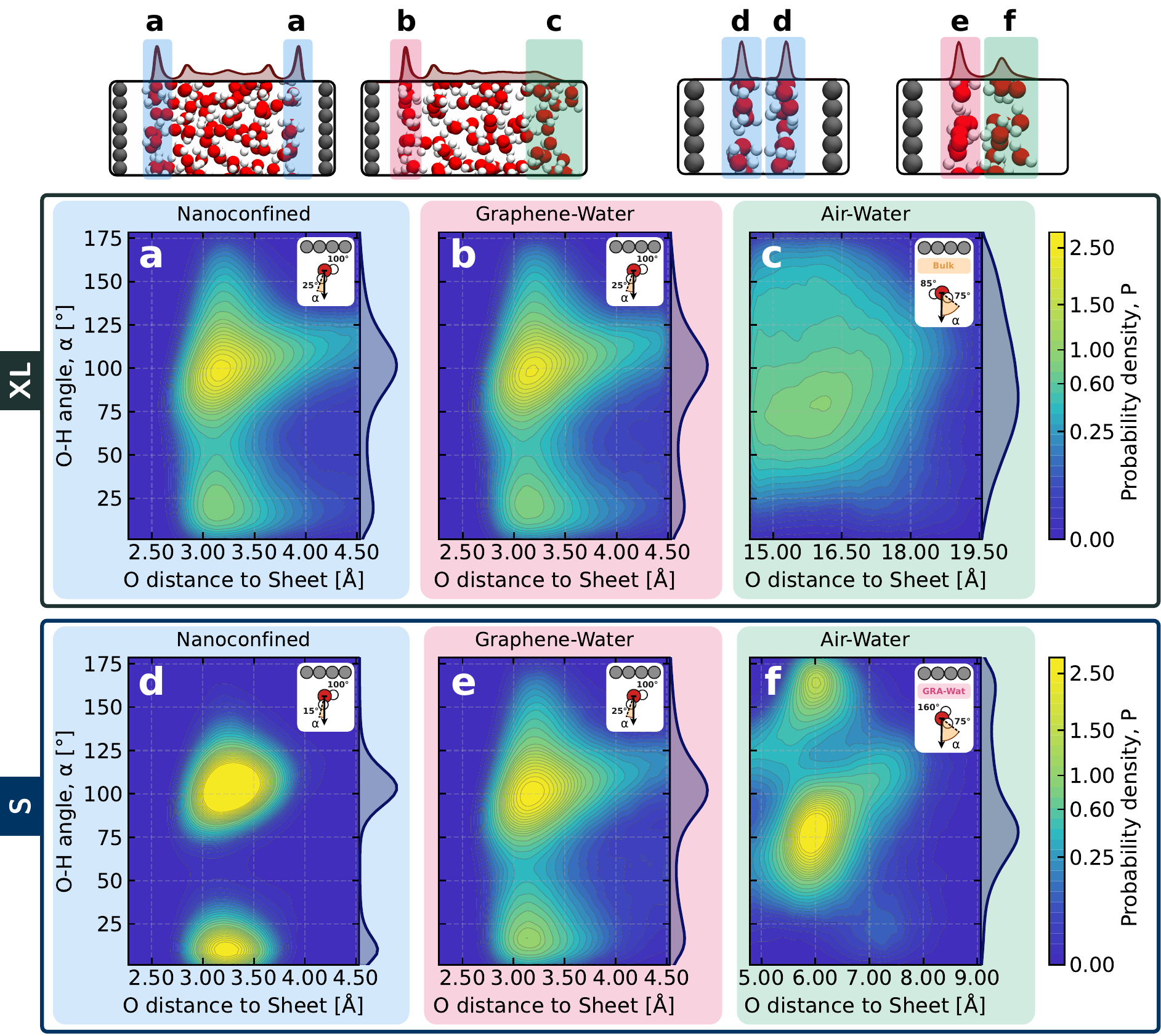}
    \caption{Interfacial orientational signatures of water at weak and strong confinement.
    Representative snapshots illustrate the systems and the definition of the interfacial regions.
    The main plots show the distributions of O–H bond orientations with respect to the surface normal as a function of the oxygen distance from the graphene sheet.
    For each system, the corresponding 1D projections shown to the right of the angle–distance maps are obtained by integrating over the distance coordinate and highlight the most probable interfacial orientations.
    The inset illustrates the angle definition: 0$^{\circ}$ corresponds to an O–H bond pointing away from the graphene sheet, whereas 180$^{\circ}$ corresponds to a bond pointing toward the graphene sheet.
    The top panel shows results for the XL systems and the bottom panel for the S system.}
    \label{fig:fig3}
\end{figure*}

Hydrogen bonding plays a central role in determining the behaviour of liquid water, so we now assess how confinement modifies the hydrogen-bond network at graphene interfaces.
To illustrate these effects, we focus on two representative cases: a weakly confined slit (XL) and a strongly confined one (S).
The accompanying snapshots in Figure~\ref{fig:fig2} highlight the interfacial regions.
For the graphene–water interface, whether in confined or interfacial systems, interfacial molecules are defined as those located between the graphene surface and the first minimum of the oxygen density profile.
At the air–water interface, interfacial molecules are taken as those located above the Gibbs dividing surface shifted by 3.1~{\AA}, which approximates the thickness of a molecular layer and is commonly used in previous studies~\cite{yuki_free_oh_2018, chiang_2022, wang_ange_2024}.
In our analysis, each water molecule is characterized by the number of hydrogen bonds it donates (D) and accepts (A), using the geometric definition of Ref.~\citenum{luzar_chandler}.
For clarity, we group molecules into the four standard hydrogen-bonding motifs DA, DDA, DAA, and DDAA.
The few molecules that fall outside these primary motifs, such as species with zero or one hydrogen bond or higher-coordination motifs, are treated following common practice in previous works~\cite{yuki_free_oh_2018, chiang_2022, wang_ange_2024} and assigned to the nearest of the four classes.

Under weak confinement (Figure~\ref{fig:fig2}, left), we see that the graphene–water interface behaves similarly in both the slit-pore and interfacial systems.
The average number of hydrogen bonds per molecule is close to 3 in both cases, with comparable intralayer (within the layer adjacent to graphene) and interlayer (between the first and second layers) contributions.
The distribution of donor–acceptor motifs is likewise nearly identical, with DDAA as the most common motif followed by DDA.
In contrast, the air–water interface shows a slight increase in the average number of hydrogen bonds, arising from differences in both intralayer and interlayer connectivity.
Unlike the graphene–water interface, the free surface does not impose solid-induced orientational constraints, allowing subsurface water to reorganise and partially compensate for missing neighbours.
Consistent with this, the air–water interface exhibits a larger fraction of DDAA species than the graphene–water interface.

The situation changes markedly under strong confinement (Figure~\ref{fig:fig2}, right).
In the S pore, the average number of hydrogen bonds at the graphene–water interface increases from approximately 2.6 in the interfacial system to about 3, and the resulting hydrogen-bonding environment differs entirely from that of the corresponding open system.
This breakdown of interfacial-like behaviour is consistent with previous observations~\cite{sergi_nanoconf_2019}, where strong deviations from standard interfacial characteristics were also reported.
In extreme cases, such as monolayer water, topological frustration in the hydrogen-bond network has even been shown to give rise to ultrafast diffusion and unusual dynamical responses, demonstrating how confinement can significantly modify hydrogen-bond connectivity and transport~\cite{acs_nanolett_2024_ultrafast}.
In the bilayer system examined here, the hydrogen-bond network is dominated by DDAA motifs, a pattern that differs substantially from the interfaces considered under weak confinement.
This prevalence of fully coordinated motifs reflects the tendency of water in angstrom-scale pores to adopt highly structured arrangements, where nearly crystalline ordering has been reported~\cite{molinero_bilayer_2010, franzese_2021}.
When the system is opened to vacuum, this ordering is lost, and the hydrogen-bond environment reverts to one characteristic of a typical interfacial system.

These results show that strong confinement does more than enhance interfacial structuring: it substantially reorganises the hydrogen-bond network.
In this regime, the graphene–water interface no longer resembles its interfacial counterpart and instead enters a distinct hydrogen-bond environment characteristic of angstrom-scale confinement.

\subsection*{Weak Confinement and Graphene-Water Share Similar Orientation Ordering}

Having examined how confinement affects the hydrogen-bond network, we now turn to a complementary descriptor of interfacial structure: the orientational ordering of water molecules.
Orientation provides an additional measure of how confining surfaces influence the arrangement of interfacial water.

Figure~\ref{fig:fig3} characterises the orientation of interfacial water by evaluating the angles formed between each O–H bond vector and the axis normal to the graphene sheet.
In this convention, 0$^{\circ}$ corresponds to an O–H bond pointing away from the graphene surface, whereas 180$^{\circ}$ indicates a bond oriented toward the graphene surface.
Under weak confinement (Figure~\ref{fig:fig3}, top), the graphene–water interface displays nearly identical orientational patterns in the slit-pore and open systems.
The distributions exhibit a pronounced maximum near 100$^{\circ}$, corresponding to O–H bonds lying largely parallel to the surface, together with a smaller feature near 25$^{\circ}$ associated with a minority of bonds pointing into the bulk.
The 1D projection to the right of the 2D map highlights these features, showing peaks at approximately 25$^{\circ}$ and 100$^{\circ}$, consistent with established orientational signatures at hydrophobic interfaces~\cite{sergi_nanoconf_2019, chris_dufils_2024}.
This observation mirrors the hydrogen-bonding results discussed earlier and reinforces that, once several water layers can form, the presence of a second graphene wall has little influence on the interfacial structure.
In contrast, the air–water interface shows a much broader orientational distribution, lacking the directional preferences imposed by a solid surface.
We note that this reflects the use of a Gibbs dividing surface to define the interface; an instantaneous interface would sharpen some of these features~\cite{wc_2010}.
Nevertheless, the overall contrast with the solid–liquid interfaces considered here remains unchanged.

Under strong confinement (Figure~\ref{fig:fig3}, bottom), the orientational ordering changes substantially.
For the bilayer slit, the distribution at the graphene–water interface becomes noticeably sharper and more structured than in the corresponding interfacial system.
In particular, it exhibits an enhanced maximum around 100$^{\circ}$ and a less prominent low-angle feature near 15$^{\circ}$, both consistent with the highly ordered arrangements identified earlier for bilayer water.
Turning to the region identified as the air–water interface, we observe a pronounced maximum near 75$^{\circ}$ together with a smaller feature around 160$^{\circ}$.
However, this should not be interpreted as a conventional vapour boundary.
In a bilayer slit, the outer layer lies in proximity to the graphene–water interface, and the orientational signal in this region therefore reflects the influence of the solid surface rather than the response of a true air–water interface.
This distinction becomes evident when comparing the orientational distributions in the weakly and strongly confined systems: only under strong confinement does the ``air–water'' region develop such structured features.
 
Together with the hydrogen-bonding analysis, these results demonstrate that strong confinement drives the system into a distinct interfacial regime, one that clearly differs from both wider pores and open graphene–water environments.

\section*{Discussion}

We investigated the solid–liquid structuring of water in nanoconfined and interfacial environments.
Our results reveal a clear transition between weak and strong confinement.
When more than three water layers can form, the graphene–water interface is essentially unaffected by the presence of a second wall and remains comparable to its open-interface counterpart.
At narrower separations, however, the available space becomes the dominant factor, and confinement drives the system into markedly more ordered arrangements, as exemplified by the bilayer case discussed above.
In this regime, the restricted geometry does more than intensify interfacial ordering; it alters the underlying structural motifs of the liquid, marking the emergence of a distinct structural regime.
Once the accessible gap falls below three layers, the graphene–water interface therefore enters a distinct nanoconfined regime, whereas wider slits retain interfacial characteristics.
The three-layer case lies on the brink of this crossover: its interfacial structure is largely recovered, although small residual deviations remain.

The separation between interfacial-like and confinement-driven behaviour is consistent with recent experimental observations~\cite{yongkang_natcoms_2025}, supporting a developing consensus on the respective roles of interface and confinement effects~\cite{chialvo_2016, sergi_nanoconf_2019, barragan_dielectric_2020, martina_dominik_2022, netz_thz_ir_2024, yongkang_natcoms_2025}.
By directly comparing nanoconfined and open graphene–water systems within a common framework, our results provide a molecular-level picture that clarifies how these regimes connect, and complements earlier studies.
Recognising this distinction is more than a semantic exercise: the ability to identify when water behaves as an interfacial liquid and when it enters a genuinely confined regime is essential for interpreting experiments and for predicting the performance of technologies that rely on nanoscale control of water.
This is particularly relevant in the context of the emerging Water Age, where filtration, energy conversion, and electrochemical technologies increasingly depend on how water reorganises at surfaces and within narrow pores.
We emphasise that these conclusions concern the structural properties analysed here, namely density layering, hydrogen bonding, and orientational ordering, under the thermodynamic conditions of our simulations.
Dynamical properties, different temperatures or pressures, chemical interactions with the confining walls, as well as defects and substrate flexibility (including ripples), may lead to additional effects not captured in the present work, and exploring these aspects remains an important direction for future study.

The broader implications of these structural differences connect to an extensive body of computational and theoretical work showing that nanoconfinement can substantially alter the behaviour of water~\cite{algara-siller_square_2015, Corsetti2016, secchi_2016_oh_prl, secchi_nat_2016,  fumagalli_dielectric_2018, volt_control_2019, know_gap_2019, franzese_2021, robin_2021, kavokine_friction_2022, vk_chris_2022, vk_chris_2022_correction,  robin_2023, quasionedimensional, chris_dufils_2024, jiang_rich_2024, Wang2025_fumagalli}.
The molecular-level insight provided here complements and extends these efforts by directly linking changes in hydrogen bonding and orientational ordering to the crossover between interfacial and confined regimes.
Adjustments to the hydrogen-bond network can influence interfacial properties such as hydrophobicity, dielectric response, and the orientation of interfacial dipoles, among others, illustrating how confinement modifies the microscopic characteristics of the interface~\cite{pezzotti_2017, pezzotti_2021_finger}.

These findings also underline that the properties of the confining walls play an important role in shaping interfacial structure.
Our analysis shows how, in the systems considered here, geometric constraints, specifically the number of molecular layers that can form, primarily determine whether water behaves in an interfacial-like or a confined regime.
Beyond geometry, the chemical character of the confining surface can also modify interfacial structure at fixed separation, as illustrated by materials such as hexagonal boron nitride, which interact more strongly with water and may enable confinement-controlled chemistry~\cite{strano_2015, versatile_mlb_2019, proton_comtet_2020, wang_ange_2024, marx_chem_2017, marx_conf_rev_2021, reactivity_2025_interfacial_or_conf}.
Mechanical and electronic tuning provide additional control. Mechanical deformation of two-dimensional materials alters interfacial interaction strengths and effective confinement geometry, influencing interfacial structure and dynamics~\cite{strain_2011, Zhang2013}, while tuning electronic structure through Moir\'e patterns or twist angles can strongly modify water friction at solid interfaces~\cite{moire_2024}.
Finally, although our analysis focuses on pure water at neutral graphene interfaces, the physical picture developed here for distinguishing interfacial and genuinely nanoconfined water applies more broadly.
In the presence of ions, this same distinction remains relevant, but additional effects such as interfacial charge, counterion accumulation, and solvent polarization can either amplify or attenuate confinement effects~\cite{perkin_2012, siria_giant_2013, perkin_2016, secchi_2016_oh_prl, fumagalli_dielectric_2018, wat_charged_2021, chris_dufils_2024, kara_pairing_2024, Wang2025_fumagalli}, highlighting the interplay of surface chemistry, mechanics, electronic structure, electrostatics, and geometric confinement, among others, in setting interfacial behaviour.

Some aspects of the structuring observed under extreme confinement may reflect the elevated pressures that naturally arise within angstrom-scale cavities~\cite{algara-siller_square_2015, vk_chris_2022, vk_chris_2022_correction, reactivity_2025_interfacial_or_conf}.
Disentangling geometric effects from the underlying thermodynamic conditions remains an active topic~\cite{scox_2022, barragan_pressure_2023, artacho_2025, reactivity_2025_interfacial_or_conf}.
Nevertheless, confinement offers a practical and well-controlled route to access high-pressure regimes that are otherwise difficult to achieve in bulk liquids.
With increasingly precise methods for fabricating nanoscale channels~\cite{radha_b_2016, geim_2019, capillary_2_2023, boya_protocol_2024, Radenovic_2024}, the combination of geometric confinement and tailored surface chemistry provides a powerful platform for probing water under conditions that are otherwise inaccessible.
More broadly, the molecular-level insight obtained here offers a coherent framework for understanding how interface and confinement effects shape water across a wide range of environments, from extended surfaces to angstrom-scale cavities.


\section*{Methods}
\textbf{Machine Learning Potential.}
The MLP used in this work was developed using the MACE architecture~\cite{mace_canonical} with 128 equivariant channels, two message-passing layers, and a 6~{\AA} cutoff.
The resulting model captures semi-local interactions through an effective receptive field of about 12~{\AA}, set by the product of the number of layers and the per-layer cutoff.
The final energy and force validation root-mean-square errors were 0.6~meV/atom and 16.6~meV/{\AA}, respectively.

The MLP was trained to reproduce the revPBE-D3(0) reference potential energy surface using the CP2K/Quickstep code~\cite{cp2k_2020}, with settings consistent with our previous work \cite{gra_wat_acid_2025, reactivity_2025_interfacial_or_conf}.
We specifically used the revPBE-D3(0)~\cite{revpbed3_1, revpbed3_2} functional due to its robust performance in reproducing the structure and dynamics of liquid water~\cite{angelos_dft_water_2016, tobias_vdw_2016, ondrej_revpbe_2017}, as well as its reliable description of both water–graphene interaction energies~\cite{Brandenburg2019} and the air–water interface~\cite{ohto_2019}.

Our model was initially trained on the data set from Ref.~\citenum{reactivity_2025_interfacial_or_conf}, which has been extensively validated to accurately describe both bulk and nanoconfined water across a broad range of conditions, including multiple slit widths.
To further refine the model for the present study, we expanded this data set to include configurations at slit widths of approximately 19 and 29~{\AA}.
Under these weakly confined conditions, we additionally incorporated graphene–water–vacuum configurations to ensure a faithful representation of the interfacial environments relevant to this work.

\textbf{Molecular Dynamics Simulations.}
All simulations were performed using the ASE software~\cite{ase_2017} with the MLP developed in this work.
The temperature was set to 300~K and maintained in the NVT ensemble using a CSVR thermostat~\cite{csvr_2007} with a 0.05~ps time constant.
The dynamics were propagated with a 1.0~fs time step using standard hydrogen masses.
In all cases, the graphene sheets were kept fully rigid throughout the simulations.
All systems were modelled in orthorhombic simulation cells with periodic boundary conditions applied in all three spatial directions.
To prevent spurious interactions between periodic replicas, we introduced a 15~{\AA} vacuum along the axis perpendicular to the interfaces, which exceeds the model’s effective receptive field.
Each system was equilibrated for 75~ps, followed by 350~ps of production sampling.
Across all simulated systems, this amounts to a total sampling time of 4.25~ns.
Such sampling would be prohibitively expensive with \textit{ab initio} molecular dynamics, highlighting the benefit of using MLPs to reach both larger system sizes and longer trajectories.
For each system, the trajectory was divided into three blocks, and all error bars represent the standard error across these blocks.

\begin{acknowledgements}

The concept of the ``Water Age'' emerged during the many stimulating and fun conversations between Mischa Bonn, Lyd\'eric Bocquet, and Angelos Michaelides during the development of the n-AQUA ERC Synergy project.
X.R.A. and A.M. acknowledge support from the European Union under the ``n-AQUA'' European Research Council project (Grant No. 101071937). 
C.S. acknowledges financial support from the Deutsche Forschungsgemeinschaft (DFG, German Research Foundation) project number 500244608, as well as from the Royal Society grant number RGS/R2/242614.
This work used the ARCHER2 UK National Supercomputing Service via the UK’s HEC Materials Chemistry Consortium, funded by EPSRC (EP/F067496).
We also utilized resources from the Cambridge Service for Data Driven Discovery (CSD3), supported by EPSRC (EP/T022159/1) and DiRAC funding, with additional access through a University of Cambridge EPSRC Core Equipment Award (EP/X034712/1).
We also acknowledge EuroHPC Joint Undertaking for awarding the project ID EHPC-REG-2024R02-130 access to Leonardo at CINECA, Italy.
\end{acknowledgements}

\section*{Data Availability}
All data required to reproduce the findings of this work will be made openly available on GitHub upon acceptance of this manuscript.

\section*{References}
%


\end{document}